\documentclass[twocolumn,showpacs,preprintnumbers,amsmath,amssymb]{revtex4}

\usepackage{graphicx}
\usepackage{dcolumn}
\usepackage{bm}

\begin{document}

\title{Walks on weighted networks}

\author{An-Cai Wu,$^{1}$ Xin-Jian Xu,$^{2}$ Zhi-Xi Wu,$^{1}$ and Ying-Hai Wang$^{1,}$\footnote{For correspondence: yhwang@lzu.edu.cn}}
\affiliation{$^{1}$Institute of Theoretical Physics, Lanzhou University, Lanzhou Gansu 730000, China\\
$^{2}$Department of Electronic Engineering, City University of
Hong Kong, Kowloon, Hong Kong, China}

\date{\today}

\begin{abstract}
We investigate the dynamics of random walks on weighted networks. Assuming that
the edge's weight and the node's strength are used as local information by a
random walker, we study two kinds of walks, weight-dependent walk and
strength-dependent walk. Exact expressions for stationary distribution and
average return time are derived and confirmed by computer simulations. We
calculate the distribution of average return time and the mean-square
displacement for two walks on the BBV networks, and find that a
weight-dependent walker can arrive at a new territory more easily than a
strength-dependent one.
\end{abstract}


\maketitle

There has been a long history of studying random walks to model
various dynamics in physical, biological, social, and economic
systems \cite {Spitzer}. A large body of theoretical results are
available for random walks performed on regular lattices and on
the Cayley (or regular) trees \cite {Barber, Hughes}, which are
defined to be trees with homogeneous vertex degree. However, it
has been suggested recently that more complex networks as opposed
to regular graphs and conventional random graphs \cite {Erdos} are
concerned to real worlds. Particularly, important classes of
random graphs such as small-world networks \cite {Watts} and
scale-free networks \cite {Barabasi} were proposed and have been
examined in the last several years. These networks share some
important properties with real networks, such as the clustering
property, short average path length, and the power-law of the
vertex degree distributions. They have been applied to the
analysis of various social, engineering, and biological networks
including epidemic spreading, percolation, and synchronization,
\emph{et al} \cite {Watts_1, Dorogovtsev, Newman, Strogatz}.

Recently, there have been several studies of random walks on
small-world networks \cite {Pandit, Lahtinen, Almaas, Parris} and
on scale-free networks \cite {Adamic, Noh, Gallos, Yang}. Most of
studies of random walks focus on unweighted networks, however, the
study of the dynamics of random walks on weighted networks is
missing while most of real networks are weighted characterized by
capacities or strengths instead of a binary state (present or
absent) \cite {Yook, Zheng, Park, Barrat, Wu, Guimera}. In the
weighted networks, a weight $w_{ij}$ is assigned to the edge
connecting the vertices $i$ and $j$, and the strength of the
vertex $i$ can be defined as
\begin{equation}
s_{i}=\sum_{j\in\nu(i)}w_{ij}, \label{str}
\end{equation}
where the sum runs over the set $\nu(i)$ of neighbors of $i$. The
strength of a node integrates the information about its
connectivity and the weights of its links.

In this paper we study the dynamics of random walk processes on
weighted networks by means of BBV model \cite {Barrat}. The model
starts from an initial number of completely connected vertices
$m_{0}$ with a same assigned weight $w_{0}$ to each link. At each
subsequent time step, addition of a new vertex $n$ with $m_{0}$
edges and corresponding modification in weights are implemented by
the following two rules: (i) The new vertex $n$ is attached at
random to a previously existing vertex $i$ with the probability
that is proportional to the strength of node $i$,
$s_{i}/\sum_{j}s_{j}$, implying new vertices connect more likely
to vertices handling larger weights. (ii) The additional induced
increase $\delta$ in strength $s_{i}$ of the $i$th vertex is
distributed among its nearest neighbors $j\in{\cal V}(i)$
according to the rule
\begin{equation}
w_{ij} \rightarrow w_{ij} + \delta \frac{w_{ij}}{s_{i}},
\label{bbv}
\end{equation}
which considers that the establishment of a new edge of weight
$w_{0}$ with the vertex $i$ induces a total increase of traffic
$\delta$ that is proportionally distributed among the edges
departing from the vertex according to their weights. The BBV
model suggests two ingredients of self-organization of weighted
networks, strength preferential attachment and weight evolving
dynamics \cite {Barrat}. Considering an arbitrary finite weighted
network which consists of nodes $i=1,\ldots,N$ and links
connecting them. The connectivity is represented by the adjacency
matrix $\textbf{A}$ whose element $a_{ij}=1$ if there is a link
from $i$ to $j$, and $a_{ij}=0$ otherwise. The information of
edges weight is represented by matrix $\textbf{W}$ whose element
$w_{ij}$ is the weight of the edge between $i$ and $j$. We
restrict ourselves to an undirected network $a_{ij}=a_{ji}$ and
symmetrical edge's weight $w_{ij}=w_{ji}$.

Assuming that edge's weight and node's strength are used as local information
by a random walker, we define two kinds of walks, weight-dependent walk and
strength-dependent walk. For weight-dependent walk, a walker chooses one of its
nearest neighbors with the probability that is proportional to the weight of
edge linked them. The transition probability from node $i$ to its neighbor $j$
is
\begin{equation}
p_{i \rightarrow j}^{w}=\frac{w_{ij}}{s_{i}}. \label{pijw}
\end{equation}
When time becomes infinite, one can find the walker staying at
node $i$ with the probability $P_{i}^{w}$, which is defined as the
stationary distribution. Following the ideas developed by Noh and
Rieger \cite {Noh}, we can write
\begin{equation}
P_{i}^{w}=\frac{s_{i}}{\sum_{j}s_{j}}, \label{piw}
\end{equation}
namely, the larger strength a node has, the more often it will be
visited by a random walker.

For strength-dependent walk, a walker at node $i$ at time $t$
selects one of its neighbors with the probability which is
proportional to the selected node's strength to which it hops at
time $t+1$. The transition probability from node $i$ to its
neighbor $j$ is
\begin{equation}
P_{i \rightarrow j}^{s}=\frac{s_{j}a_{ij}}{s_{i}^{'}},
\label{pijs}
\end{equation}
where $s_{i}^{'}=\sum_{l\in\nu(i)}s_{l}$ and $\nu(i)$ denotes the
set of all neighboring vertices of node $i$. Supposing that a
walker starts at node $i$, and the probability that the walker at
node $k$ after $t$ time steps denoted by $P_{ik}^{s}(t)$, then the
master equation for the probability $P_{i \rightarrow j}^{s}$ to
find the walker at node $j$ at time $t+1$ is
\begin{equation}
P_{i \rightarrow j}^{s}(t+1)=
\sum_{k}\frac{s_{j}a_{kj}}{s_{k}^{'}}P_{ik}^{s}(t). \label{pmq}
\end{equation}
An explicit expression for the transition probability $P_{i
\rightarrow j}^{s} (t)$ to go from node $i$ to node $j$ in $t$
steps follows by iterating Eq. (\ref{pmq})
\begin{equation}
P_{i \rightarrow j}^{s}(t)=
\sum_{j_{1},\ldots,j_{t-1}}\frac{s_{j_{1}}a_{ij_{1}}}{s_{i}^{'}}\cdot\frac{s_{j_{2}}a_{j_{1}j_{2}}}{s_{j_{1}}^{'}}
\cdots\frac{s_{j}a_{j_{t-1}j}}{s_{j_{t-1}}^{'}}. \label{ipmq}
\end{equation}
Comparing the expressions for $P_{i \rightarrow j}^{s}(t)$ and $P_{j
\rightarrow i}^{s}(t)$, we get
\begin{equation}
s_{i}s_{i}^{'}P_{i \rightarrow j}^{s}(t)=s_{j}s_{j}^{'}P_{j
\rightarrow i}^{s}(t). \label{comp}
\end{equation}
This is a direct consequence of the undirectedness of the network.
We can also define the probability $P^{s}_{i}$ as the stationary
distribution when the evolving time becomes infinite. Eq.
(\ref{comp}) implies that $s_{i}s_{i}^{'}P_{j}^{s}=s_{j}
s_{j}^{'}P_{i}^{s}$, and therefore one can obtain \cite {Noh}
\begin{equation}
P_{i}^{s} =\frac{s_{i}s_{i}^{'}}{\sum_{l}s_{l}s_{l}^{'}}.
\label{pis}
\end{equation}

Now we discuss the stationary distribution for edges, i.e., the
probability that an edge is chosen by the walker to follow as the
evolving time becomes infinite. In unweighted networks, all edges
are equal and a random walker will choose one of its neighboring
edges at the same probability. So each edge in the network has the
same probability to be chosen by the walker when $t \rightarrow
\infty$ for weight-dependent walk. In weighted networks, however,
the walker will choose an edge according to the weight of it or
the strength of the node connected by it. Then the relation
between the stationary distribution for edges $P_{e_{ij}}$ and the
stationary distribution for nodes $P_{i}$ can be written as
\begin{equation}
P_{e_{ij}}=P_{i}P_{i \rightarrow j}+P_{j}P_{j \rightarrow i},
\label{rel}
\end{equation}
where $P_{i \rightarrow j}$ is the transition probability from
node i to node j. For weight-dependent walk, substituting Eqs.
(\ref{pijw}) and (\ref{piw}) into Eq. (\ref{rel}), we obtain the
stationary distribution for edges
\begin{equation}
P_{e_{ij}}^{w} = \frac{2 w_{ij}} {\sum_{k,l} w_{kl}}.
\label{peijw}
\end{equation}
For strength-dependent walk, substituting Eqs. (\ref{pijs}) and
(\ref{pis}) into Eq. (\ref{rel}), we obtain the stationary
distribution for edges
\begin{equation}
P^{s}_{e_{ij}} = \frac{2 s_{i}s_{j}}{\sum_{k,l}s_{k}s_{l}a_{kl}}.
\label{peijs}
\end{equation}

\begin{figure}
\includegraphics*[width=\columnwidth]{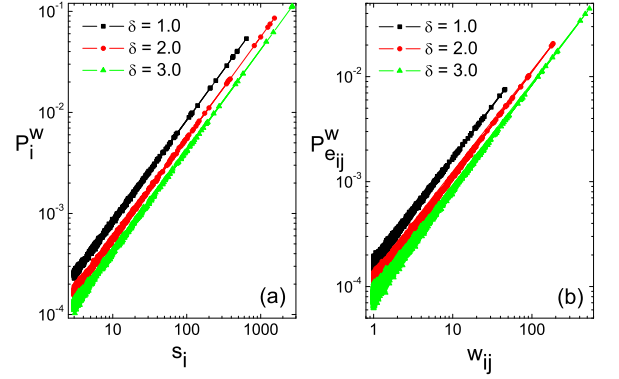}

\caption{Log-log plots of $P_{i}^{w}$ vs. $s_{i}$ (a) and
$P^{w}_{e_{ij}}$ vs. $w_{ij}$ (b) in the BBV network with
$N=1000$, $m_{0}=3$ and different values of $\delta$. Slopes of
all the curves are equal to $0.9999 \pm 0.0006$. The data were
obtained after walking $10^{8}$ steps on the network.}\label{fig1}
\end{figure}

\begin{figure}
\includegraphics*[width=\columnwidth]{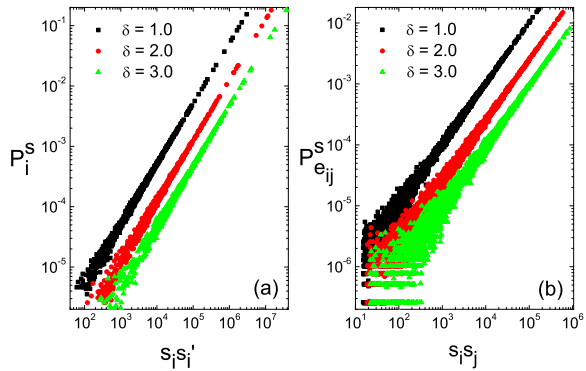}

\caption{Log-log plots of $P_{i}^{s}$ vs. $s_{i}s_{i}^{'}$ (a) and
$P^{s}_{e_{ij}}$ vs. $s_{i}s_{j}$ (b) in the BBV network with
$N=1000$, $m_{0}=3$ and different values of $\delta$. Slopes of
all the curves are equal to $1.018 \pm 0.003$. The data were
obtained after walking $10^{8}$ steps on the network.}\label{fig2}
\end{figure}

In Fig. \ref{fig1} we plot $P_{i}^{w}$ vs. $s_{i}$ (Fig.
\ref{fig1}(a)) and $P^{w}_{e_{ij}}$ vs. $w_{ij}$ (Fig.
\ref{fig1}(b)) in log-log scale in the BBV network. The power-law
property of Eqs. (\ref{piw}) and (\ref{peijw}) is presented in
excellent agreement with the numerical results. In Fig. \ref{fig2}
we show the log-log plots of $P_{i}^{s}$ vs. $s_{i}s_{i}^{'}$
(Fig. \ref{fig2}(a)) and $P^{s}_{e_{ij}}$ vs. $s_{i}s_{j}$ (Fig.
\ref{fig2}(b)) in the BBV model. The numerical results are also in
good agreement with Eqs. (\ref{pis}) and (\ref{peijs}).

Next we study average return time which is the average time spent by a walker
to return to its origin. From its definition, we can easily obtain that the
average return time is equal to the reciprocal of the stationary distribution.
The average return time for node $i$ is
\begin{equation}
\langle T_{ii}^{w} \rangle = \frac{1}{P_{i}^{w}} =
\frac{\sum_{j}{s_{j}}}{s_{i}}, \label{tiiw}
\end{equation}
for weight-dependent walk and
\begin{equation}
\langle T^{s}_{ii} \rangle = \frac{1}{P_{i}^{s}}
 = \frac{\sum_{j}s_{j}s_{j}^{'}}{s_{i}s_{i}^{'}}. \label{tiis}
\end{equation}
for strength-dependent walk, respectively.

Using the same methods, the average return time for edge $e_{ij}$
can also be obtained
\begin{equation}
\langle T^{w}_{e_{ij}} \rangle = \frac{\sum_{k,l}w_{kl}}{2w_{ij}},
\label{teijw}
\end{equation}
for weight-dependent walk and
\begin{equation}
\langle T^{s}_{e_{ij}} \rangle =
\frac{\sum_{k,l}s_{k}s_{l}a_{kl}}{2s_{i}s_{j}}. \label{teijs}
\end{equation}
for strength-dependent walk, respectively.

\begin{figure}
\includegraphics[width=\columnwidth]{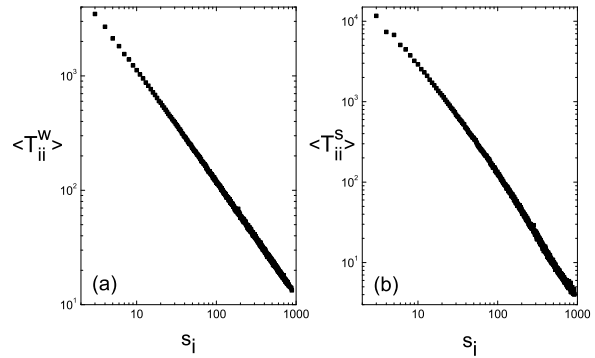}

\caption{Numerical results for the average return time of node for
the two different walks , weight-dependent walk (a) and
strength-dependent walk (b), on the BBV network with $N=1000$,
$m_{0}=3$ and $\delta=1.0$. The data were averaged on 1000
networks and obtained after walking $10^{6}$ steps in each
network.}\label{fig3}
\end{figure}

In Fig. \ref{fig3} we show the log-log plots of $\langle T_{ii}^{w} \rangle$
vs. $s(i)$ and $\langle T_{ii}^{s} \rangle$ vs. $s(i)$ in the BBV networks. The
slope of the curve in Fig. \ref{fig3}(a) is $-0.9903 \pm 0.0007$, consistent
with Eq. (\ref{tiiw}). The BBV networks have two properties in \cite {Barrat}:
(i) node's strength is proportional to node's degree, $s_{i} \sim k_{i}$; (ii)
the average nearest neighbor degree is $k_{nn}(k) \sim k^{-2+1/\beta}$ with
$\beta=(2\delta+1)/(2\delta+2)$. Considering these two ingredients, one can
observe that $s_{i}s_{i}^{'} \sim s_{i}^{1/\beta}$, and obtain $\langle
T_{ii}^{s} \rangle \sim s^{-1/\beta}_{i}$. Fig. \ref{fig3}(b) gives that
$-1/\beta=-1.355 \pm 0.006$ which agrees with the theoretical value
$-1/\beta=-(2\delta+2)/(2\delta+1)=-4/3$. In Fig. \ref{fig3} the slope of the
curve for strength-dependent walk is steeper than that for weight-dependent
walk, giving rise to a broader distribution of average return time for
strength-dependent walk. In order to confirm this point, we derive the
expression for the distribution of average return time. In BBV network, we know
that the strength distribution behaves as $P(S)\sim S^{-\gamma}$, where
$\gamma=(3+4\delta)/(1+2\delta)$ \cite{Barrat}. According to the above
analysis, we can obtain that the distribution of average return time is
$P(T^{w})\sim(T^{w})^{-1/(1+\delta)}$ for the weight-dependent walk, and
$P(T^{s})\sim1.0$ for the strength-dependent case. Thus, we can see, for those
nodes with large strength the strength-dependent walker spends more time in
visiting them than that the weight-dependent walker does. This point can be
reflected by the difference of the mean-square displacement for two walks which
is shown in the following.

\begin{figure}
\includegraphics[width=7cm]{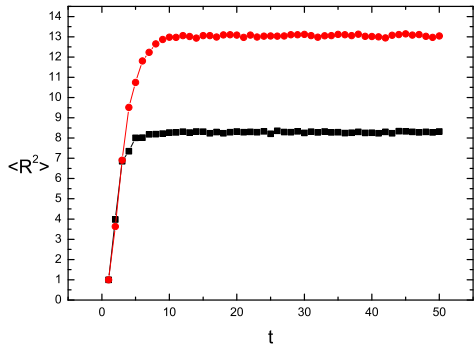}
\caption{Mean-square displacement $\langle R^{2} \rangle$ as a
function of time $t$ for two kinds of walks, weight-dependent walk
(closed circles) and strength-dependent walk (closed squares), on
the BBV network with $N=10^{5}$, $m_{0}=3$, and $\delta=1.0$. All
the plots were averaged over $1000$ different initial positions of
the walker and $20$ realizations of the network.}\label{fig4}
\end{figure}

The mean-square displacement $\langle R^{2} \rangle$ is a measure of the
distance $R$ covered by a typical random walker after performing $t$ steps
\cite {Almaas, Gallos}. To calculate this quantity we first, at each time step,
find the minimal distance from the current position of the walker to the origin
(i.e., the smallest number of steps needed for the walker to reach the origin )
using a breadth-first search method. Then we allow the walker to move through
the network until $<R^{2}>$ has saturated. Finally, we average over different
initial positions of the walker and realizations of the network. Fig.
\ref{fig4} shows the results of $\langle R^{2} \rangle$ as a function of time
$t$. We note that $\langle R^{2} \rangle$ equilibrates after a few steps to a
constant displacement value. This is a simple manifestation of the very small
diameter of the BBV network. Note also that the value of plateau is higher for
weight-dependent walk than that for strength-dependent walk. That is, a
strength-dependent walker spends more time in visiting these nodes with large
strength or large degree, i.e. a weight-dependent walker can arrive at a new
territory more easily than a strength-dependent walker.

In summary, we have studied two different types of walks on weighted networks,
weight-dependent walk and strength-dependent walk. We derived exact expressions
for the stationary distribution and the average return time for the two walk
processes, and confirmed them by simulations on BBV networks. Then we analysed
the distribution of average time for two walks and found that it is broader for
dependent-strength walk than that for dependent-weight walk on the BBV network.
Finally we computed the mean-square displacement $\langle R^{2} \rangle$. For
both walks, $\langle R^{2} \rangle$ was found to reach the saturation after a
few time steps which is a result of the very small diameter of the underlying
graph. Furthermore, the difference of average-square displacement for two walks
implies that a weight-dependent walker can arrive at a new territory more
easily than a strength-dependent walker on the BBV network.

\end{document}